\documentclass[twocolumn,showpacs,preprintnumbers,amsmath,amssymb]{revtex4}
\usepackage{graphicx}
\usepackage{bm}
\begin{document}
\title{On the evaluation of the ac-Stark shift and Kramers-Heisenberg dispersion formula}%
\author{R. Radhakrishnan}
\email{rkrishna@cusat.ac.in}
\affiliation{Department of Theoretical Physics, 
Guindy Campus, University of Madras, Chennai-25, India.}
\author{Ramesh Babu Thayyullathil}%
\affiliation{Department of physics, Cochin university of Science and
Technology, Cochin 682022, India.}
\date{\today}%
\begin{abstract}
We describe a unified approach for the determination of ac-Stark
correction and Kramers-Heisenberg dispersion formula. In both cases
the contribution from infinite intermediate summation appearing in
the expression for the corresponding matrix elements are evaluated
exactly in the dipole approximation for the ground state of hydrogen
atom using a variation of the Dalgarno-Lewis method. The analytical
expressions obtained can be efficiently used for the numerical
evaluation of matrix element for all values of incident photon
energy and comparison is made with results obtained by different
methods.
\end{abstract}
\maketitle
\section{Introduction}
\label{sect:intro}%
In the presence of intense laser field, atomic or molecular levels
are displaced (shifted or broadened) and these stimulated radiative
corrections are alternatively known as ac-Stark shift, dynamic Stark
effect  or light shift \cite{cohen}. Being a fundamental process
ac-stark effect is very well studied theoretically as well as
experimentally in wide area of atomic and molecular physics. This
effect is also very important in the area of laser trapping and
cooling and is a basic mechanism behind many nonlinear optical
effects \cite{bec}. It is a promising tool for the field of optical
communications (Optical switching methods) and  plays a very
decisive role in many current research areas
\cite{dyn,schu,glib,shin,zou,sch,rud,bach,apa,ago,chen,yak}. In this
work we have described an efficient  method to obtain a closed form
analytical expression for two of the basic process viz. ac stark
shift and elastic scattering cross section of photons for the atomic
hydrogen in the  ground state.
Our aim is to get an analytical expression for dipole dynamical
polarizability $\tau^{(2)}(\omega)$ from which one can calculate
the level shift $\delta E^{(2)}$ and level width $\Gamma^{(2)}$.
The hydrogen atom being the simplest of all atomic systems, plays
a very special role in this respect because it has  closed-form
wavefunctions. Since these wave functions are analytically known,
both the dipole dynamic polarizability and Kramer's-Heisenberg
dispersion formula for hydrogen atom can be written in a closed form.

The level shift for atomic hydrogen was previously
calculated by both perturbative and nonperturbative methods
\cite{pan,maq:83,kar:04,bak,chu:coo,gon:89,maq:90,sha:tan, ben:kri,flo:pat}.
In the perturbative method, evaluation of the higher order
matrix element is the key problem and it is achieved through
Greens function formalism \cite{pan,maq:83,kar:04} or by
solving a system of second-order differential equations
\cite{gon:89}.

The Kramers-Heisenberg formula alternatively known as
dispersion formula plays a very important role in the determination of
scattering cross-sections \cite{hei}.
Considering its importance as an age-old relation
there were only very few attempts to evaluate it
analytically \cite{mit,gav}. There were some generalization
of it and alternate forms are also available in the
literature \cite{hea:woo,hol,bre,tul}.
Recently a low-energy expansion of the Kramers-Heisenberg
formula for Rayleigh scattering process in atomic hydrogen was reported \cite{lee}.
Similar to the ac Stark shift this also involves the
evaluation of the second order matrix element.

The exact calculation of higher order process
in the perturbative formulation is nontrivial  because of the presence
of infinite  summation over intermediate states in the expression
for the  higher order matrix element. In our approach the matrix element
containing infinite summation over the whole hydrogenic
spectrum (discrete and continuum) is performed by using an
implicit summation technique due to Dalgarno Lewis \cite{dar:lew},
which  reduces the evaluation of the infinite summation
to finding a solution of some inhomogeneous differential
equation. The closed form expression which we have obtained
is  very simple and also very convenient for analytical
continuation. Thus with this method, we can very easily obtain
the relevant matrix element for radiation with energy larger
than ionization energy  (above threshold ionization),
while other methods need some kind of
approximation like Pade approximation in the
case of Coulomb Greens function formalism.
\section{Radiative correction in the ground state}
The level shift depends on the intensity $I$ and the 
frequency $\omega$ of the radiation and the complex second order shift
$\Delta^{(2)}$ is given by \cite{maq:83,kar:04,chu:coo,gon:89}
\begin{equation}%
\label{delta}%
\Delta^{(2)}(\omega)=\delta E^{(2)}-i \Gamma^{(2)}%
 =-\frac I{I_0}\tau^{(2)}(\omega)
\end{equation}%
where the real part $\delta E^{(2)}$ is the energy shift and the
imaginary part $\Gamma^{(2)}$ gives the level width and
$I_0= 7.016\times 10^{16}$ \,W/cm$^2$ is the characteristic atomic field
strength intensity.
Here $\tau^{(2)}(\omega)$ is the dipole dynamic polarizability
and in the atomic unit it can be written as
\begin{equation}
\tau ^{(2)}=\sum_n \left\{\frac{\langle g\mid \epsilon ^*\cdot
{\bf r}\mid n\rangle \langle n\mid \epsilon\cdot {\bf r}\mid
g\rangle } {(E_g-E_n+\omega )}+\frac{\langle g\mid \epsilon \cdot
{\bf r}\mid n\rangle \langle n\mid \epsilon ^*\cdot {\bf r}\mid
g\rangle } {(E_g-E_n-\omega )} \right\}%
\label{mat}%
\end{equation}%
where $E_g$ is the atomic ground state energy
and $\epsilon$ is
the polarization of the radiation. First term in the bracket is
the absorption-emission term and the second is the
emission-absorption term. The infinite summation over the complete
set of intermediate state  in equation (\ref{mat}) can be performed
exactly by defining a set of operators $F$ and $\widetilde{F}$
such that
\begin{equation}\label{opf1}
\mathbf{\epsilon}\cdot {\bf r}\left| g\right \rangle =%
\left( FH_0-H_0F+\omega F\right) \left| g\right\rangle%
\end{equation} %
\begin{equation}
\mathbf{ \epsilon}^*\cdot {\bf r}\left| g\right \rangle =\left(
\widetilde{F}H_0-H_0\widetilde{F}-\omega \widetilde{F}\right)
\left| g\right\rangle%
\label{opf2}
\end{equation}
where $H_0=-\nabla^2/2-1/r $ is the unperturbed Hamiltonian for
atomic hydrogen and  $|g\rangle=e^{-r}/\sqrt{\pi}$ is the ground
state wave function in atomic units. With these definitions and 
the closure relation
$\sum_n\mid n\rangle \langle n\mid\ =\hat{I}$ the expression in
equation (\ref{mat}) for dipole dynamical polarizability will be reduced to
\begin{equation}
\tau^{(2)}(\omega)=
\langle g\mid \mathbf{\epsilon}^*\cdot {\bf r}F\mid
g\rangle +\langle g\mid \mathbf{\epsilon}\cdot{\bf
r}\widetilde{F}\mid g\rangle.%
\label{rmat}%
\end{equation}
Thus the infinite summation over the intermediate states are
reduced to the determination of the operator $F$ and
$\widetilde{F}$. Using co-ordinate space representation and writing
$F=\mathbf{\epsilon}\cdot \mathbf{r}f(r)$ and
$\widetilde{F}=\mathbf{\epsilon}^*\cdot
\mathbf{r}\widetilde{f}(r)$,
the equations (\ref{opf1}) and (\ref{opf2}) become
\begin{eqnarray}
r\frac{d^2}{dr^2}f(r)+(4-2r)\frac d{dr}f(r)+(2\omega r-2)f(r)=2r
\\ \label{f12}
r\frac{d^2}{dr^2}\widetilde{f}(r)+(4-2r)\frac
d{dr}\widetilde{f}(r)-(2\omega r+2)\widetilde{f}(r)=2r%
 \label{f13}%
\end{eqnarray}
By the method of Laplace transform \cite{inc} for the
solution of differential equation we can obtain the solutions to
the above differential equations as
\begin{equation}
\label{f1} 
f(r)=\frac {1}{\omega} -\frac {1}{2\omega
^3}\Phi(1,1,\lambda,r)
\end{equation}and
\begin{equation}
\label{f2}
 \widetilde{f}(r)=\frac {1}{\omega} +\frac {1}
{2\omega ^3}\widetilde{\Phi}(1,1,\widetilde{\lambda},r)
\end{equation}
with
\begin{equation}
\label{phi} 
\Phi(p,q,\lambda,r)=  \int \limits_{\lambda} ^1
ds\,e^{-r(s-1)}K(p,q,\lambda,s)
\end{equation}
\begin{equation}
\label{psi}
\widetilde{\Phi}(p,q,\lambda,r)=\int \limits_1^{\lambda}
ds\,e^{-r(s-1)}\widetilde{K}(p,q,\lambda,s)
\end{equation}
\begin{equation}
\label{ker} %
K(p,q,\lambda,s)=\left(
\frac{1-\lambda}{1+\lambda}\right )^{\frac 1{\lambda}}%
\left( s+\lambda \right) ^{p+\frac {1}{\lambda}}%
\left( s-\lambda\right) ^{q-\frac {1}{\lambda}}
\end{equation}
\begin{equation}%
\label{ker2}%
\widetilde{K}(p,q,\lambda,s)=\left(\frac{\lambda-1}{\lambda+1}\right)%
^{\frac 1{\lambda}} \left( \lambda+s \right) ^{p+\frac {1}{\lambda}}%
\left(\lambda-s\right) ^{q-\frac {1}{\lambda}}
\end{equation}
where $\lambda=\sqrt{1-2\omega}$ and
$\widetilde{\lambda}=\sqrt{1+2\omega}$. In this work we obtain the solutions
with integer values of $p$ and $q$. But in general for the purpose of analytic
continuation it can be complex and $\lambda$ also can become complex depending
on the value of the frequency $\omega$.
Now using equations (\ref{phi}) and (\ref{psi})
in equation (\ref{rmat}) the final form of the dipolar polarizability
becomes
\begin{equation}%
\label{ftau}
\tau^{(2)}(\omega)=\frac2 {3\omega^3} \int\limits_0^\infty
dr\,e^{-2r}\,r^4\{\widetilde{\Phi}(1,1,\widetilde{\lambda},r)
-\Phi(1,1,\lambda,r)\}
\end{equation}%
The limit $\omega\rightarrow 0$  of the expression for $\tau^{(2)}(\omega)$
can be shown to approach the value corresponding to the static dipolar
polarizability which is $9/2$.
\section{Elastic scattering of photons}
\label{sec:scat}
The differential scattering cross section for low energy
elastic scattering of photons with frequency $\omega$,
by atoms is given by the dispersion formula \cite{hei,mit,gav}%
\begin{equation}
\frac{d\sigma }{d\Omega}=%
a_0^2\left(\epsilon\cdot\epsilon'\right)^2 %
\mid M(\omega) \mid ^2= a_0^2\left(\epsilon\cdot\epsilon'\right)^2%
\mid 1-P(\omega)-P(-\omega)\mid ^2 ~\label{dsred}%
\end{equation}
where $a_0$ is the Bohr radius, $M$ is the Kramers-Heisenberg
matrix element, $\epsilon$ and $\epsilon'$ respectively are the
initial and final polarization  of photons and  $P(\omega)$ in atomic
units is given as \cite{gav}
\begin{equation}
P(\omega)=-\frac 23\sum_n \frac{\langle g\mid \mathbf{p}\mid n\rangle
\cdot \langle n\mid \mathbf{p}\mid g\rangle}{E_g-E_n+\omega }. %
\label{pk}
\end{equation}
Here $\mathbf{p}$ is the momentum
operator 
and the summation is over the complete set of states including continuum
states. It is useful to note from equation (\ref{dsred}) that the
differential cross section for coherent scattering of photons is
just the Thompson cross section modified by the dynamic polarizability.
In a similar fashion we can consider Raman scattering where
initial and final states are different. The analytical expressions
for $P(\omega)$ were derived earlier using Schwartz and Teinman method
\cite{mit} and Coulomb Greens function (CGF) formalism \cite{gav}.

Using a slight variation of the formalism in the previous section
we can easily calculate $P(\omega)$. In this case the infinite
summation over the intermediate state in equation~(\ref{pk}) is
performed  by defining a set of operators $\mathbf{U}$ and
${\widetilde{\mathbf{U}}}$ such that,
\begin{eqnarray}\label{op12}
\mathbf{p}\left|g\right \rangle =\left(
\mathbf{U}H_0-H_0\mathbf{U}+\omega
\mathbf{U}\right) \left|
g\right\rangle\\  \label{op13}
\mathbf{p}\left| g\right \rangle =\left(
\widetilde{\mathbf{U}}H_0-H_0\widetilde{\mathbf{U}}%
-\omega \mathbf{\widetilde{U}}\right) \left| g\right\rangle
\end{eqnarray}%
and the expression for $M$ will become
\begin{equation}%
M(\omega)= 1+\frac{2}3\left(\langle g\mid \mathbf{p}\cdot
\mathbf{U}\mid g\rangle +\langle g\mid\mathbf{p}\cdot
\mathbf{\widetilde{U}}\mid g\rangle\right)
 \label{rkhmat}
\end{equation}
Now the evaluation of the infinite summation is reduced to the
evaluation of the operators $\mathbf{U}$ and
$\widetilde{\mathbf{U}}$. To obtain a similar expression as in
the previous section, instead of going to momentum space
representation \cite{mit,gav},  we use the coordinate space
representation of $\mathbf{U}$ and $\widetilde{\mathbf{U}}$.
This is done by taking
$\mathbf{U}=\mathbf{r}u(r)$ and
$\widetilde{\mathbf{U}}=\mathbf{r}\widetilde{u}(r)$ and the
equations (\ref{op12}) and (\ref{op13}) become
\begin{eqnarray}
 r\frac{d^2}{dr^2}u(r)+(4-2r)\frac d{dr}u(r)+(2\omega r-2)u(r)=2i
\\ \label{j12}
 r\frac{d^2}{dr^2}\widetilde{u}(r)+(4-2r)\frac d{dr}
 \widetilde{u}(r)-(2\omega r+2)\widetilde{u}(r)=2i%
\label{j13}
\end{eqnarray}
These differential equations has the same form which appeared in the
previous section and the solutions can be written as
\begin{equation}
 \label{j1}
 u(r)= -\frac {i}{2\omega ^2}\Phi(1,1,\lambda,r)
\end{equation}
and
\begin{equation}
 \label{j2}
 \widetilde{u}(r)= \frac {i}{2\omega^2}
 \widetilde{\Phi}(1,1,\widetilde{\lambda},r)
\end{equation}
Using this in the definition of $U$ and $\widetilde{U}$ and substituting it
in equation~(\ref{rkhmat})
we get a closed form expression for the Kramers-Heisenberg matrix
element. It now takes the form
\begin{equation}
 M(\omega)= 1-\frac 4{3\omega^2}\int\limits_0^\infty
 dr\,e^{-2r}\,r^3\left\{\Phi(1,1,\lambda,r)
 -\widetilde{\Phi}(1,1,\widetilde{\lambda},r)\right\}
 \label{fkhmat}
\end{equation}
If the incident photons are unpolarized and the polarization of
the scattered photons are not observed the differential scattering
cross section will take the standard form
\begin{equation}
  d\sigma=r_0^2\frac12(1+cos^2\theta)|M(\omega)|^2d\Omega
  \label{unpol}
\end{equation}
%
\section{Discussion and Conclusion}
\label{sec:num}%
The radial integrals in equations (\ref{ftau}) and (\ref{fkhmat}) can be
done exactly \cite{isg}.  For the numerical evaluation of various
integrals it is very convenient to define the following
\begin{eqnarray}%
\label{stn}
I(p,q,\lambda,n)&=&\int\limits_0^\infty
dr\,e^{-2r}r^n\, \Phi(p,q,\lambda,r)\\ %
&=&n!\left( \frac{1-\lambda}{1+\lambda}\right )^{\frac
1{\lambda}}\int\limits_\lambda^1 ds \frac{\left( s+\lambda \right)
^{p+\frac {1}{\lambda}}\nonumber%
\left( s-\lambda\right) ^{q-\frac {1}{\lambda}}}{(1+s)^{n+1}}
\end{eqnarray}%
and
\begin{eqnarray}%
\label{stn2}
\widetilde{I}(p,q,\lambda,n)&=&\int\limits_0^\infty
dr\,e^{-2r}r^n\, \widetilde{\Phi}(p,q,\lambda,r)\\ %
&=&n!\left( \frac{\lambda-1}{\lambda+1}\right )^{\frac
1{\lambda}}\int\limits_1^\lambda ds  \frac{\left( \lambda
+s\right) ^{p+\frac {1}{\lambda}}\nonumber%
\left( \lambda-s\right) ^{q-\frac {1}{\lambda}}}{(1+s)^{n+1}}.
\end{eqnarray}
With these, the Kramers-Heisenberg matrix element and the dipole
dynamic polarizability respectively will take the following forms
\begin{equation}
 \label{fkhm}
 M(\omega)=1-\frac 4{3\omega^2}\left\{I(1,1,\lambda,3)-
 \widetilde{I}(1,1,\widetilde{\lambda},3)\right\}
\end{equation}
\begin{equation}
\tau^{(2)}(\omega) =\frac2{3\omega^3}\left\{\widetilde{I}
(1,1,\widetilde{\lambda},4)-I(1,1,\lambda,4) \right\}%
 \label{fdpl}
\end{equation} %
This is the final  form of the analytic expression for the matrix elements.
We did not attempt to rewrite these in terms of well known special
functions because the remaining integrals in equations (\ref{stn}) and (\ref{stn2})
can be done very efficiently by a simple numerical integration.

The integrals in
equation (\ref{phi}) is defined only for Re$(q-1/\lambda)>-1$. But in order to
calculate the matrix elements in equations (\ref{fkhm}) and (\ref{fdpl}) for
all values of $\omega$,  we require an analytical continuation of the
definition of $\Phi(.)$ given in equation (\ref{phi}).
It can be achieved by performing a partial integration in the definition given
in equation (\ref{phi}) and this gives
\begin{eqnarray} \label{rec}%
\Phi(p,q,\lambda,r)&=&\frac 1{q+1-\frac{1}{\lambda}}\left[K(p,q+1,\lambda)
+r\Phi(p,q+1,\lambda,r)\right.\\ \nonumber
&&\left.+\left(p+\frac{1}{\lambda}\right) \Phi(p-1,q+1,\lambda,r)\right].
\end{eqnarray}%
This is a very important recurrence relation because (\ref{rec}) can be used to analytically
continue the definition of these function for Re$(q-1/\lambda)>-2$.
We can also repeat this procedure to analytically continue the definition
to other values of Re$(q-1/\lambda)$. And for the
numerical evaluation of various integrals as a function of
radiation frequency $\omega$ we have extensively used this
relation. It is also very important to note the appearance of
simple poles in the definition of $\Phi(.)$ when $q=-1+1/\lambda$.
Since $\lambda=\sqrt{1-2\omega}$, this is due to the presence of
intermediate resonances (when $2\omega=1-1/n^2$, where
$n=2,3,\ldots$) in the expression for second order matrix element
in the perturbation theory. This also leads to a  strong frequency
dependence  of ac stark effect and scattering cross section.

Similarly for above one photon ionization threshold 
i.e. $\omega>1/2$, a simple analytic continuation makes $\lambda$
to be a purely imaginary number. Thus by this simple method, with the same
analytic expression for the matrix element, we can numerically
evaluate it for the whole physical range of photon frequency $\omega$.
Displayed in table \ref{tab:belo-pol}  are
$\tau^{(2)}(\omega)$ as a function of photon frequency in atomic units and are
compared  with values obtained by other methods.
It is purely real for $\omega<1/2$.
For $\omega>1/2$ one photon ionization is possible and this makes the
intermediate virtual state to lie in the continuum and $\tau ^{(2)}(\omega)$
becomes complex.

The dipole dynamic polarizability for photon
energy below and above one photon ionization threshold  
is given in table \ref{tab:belo-pol}. Our values are compared
with values obtained by Arnous et al. \cite{maq:83} using Coulomb Greens function
method, but with an additional factor of $\omega^{-2}$
in their result. It is very easy to see that  this factor is
missing from their result by taking the $\omega \rightarrow 0$ limit.
In this limit our result approaches the dc polarizability, which is 9/2.
The values given in  \cite{maq:83} also approaches to the same limit
provided their results are multiplied by an overall factor of $\omega^{-2}$.
Using the analytic continuation described in equation (\ref{rec}) we can calculate
$\tau ^{(2)}(\omega)$ for $\omega$ very close to intermediate resonance also.
Our result also agrees with the second order
level shift and width reported by Pan et al. \cite{pan}.
%
From the values given in tables \ref{tab:belo-pol} and \ref{tab:belo-khm}
it is useful to note the change
in sign of the level shift when the photon frequency cross these resonance
values. The values of  Kramers-Heisenberg matrix element are given in tables
\ref{tab:belo-khm}  and \ref{tab:abo-khm}. They are in good agreement
with the  results of  Gavrila \cite{gav}.
\begin{acknowledgments}Authors acknowledge the support from UGC
through DSA-COSIST Scheme
\end{acknowledgments}

\begin{table*}
\caption{\label{tab:belo-pol}Comparison of the values of dipole
dynamical  polarizability in atomic units for both below and above
one photon ionization threshold $\omega=0.5$. In the limit
$\omega\rightarrow 0$, $\tau^{(2)}(\omega)$ approaches the dc
polarizability, which is -4.5.}
\begin{ruledtabular}%
\begin{tabular}{cccccccc}
&&&&\multicolumn{2}{c}{Re $\tau^{(2)}(\omega)$}
    &\multicolumn{2}{c}{Im  $\tau^{(2)}(\omega)$}\\
$\omega$(a.u) & This work %
&ABM \footnote{Values obtained using
Coulomb Green function \cite{maq:83} by Arnous et al. with an
additional $\omega^{-2}$ factor as explained in  section
\ref{sec:num}.}%
&$\omega$(a.u) & This work & ABM \footnotemark[1]& This work & ABM\footnotemark[1]\\
\hline
0.001   &\-4.50003 & --    & 0.6  & 3.297  & --    &2.505   & --  \\
0.002   &\-4.50011 & --    & 0.7  & 2.493  & --    &1.408   & --  \\
0.02    & \-4.51066&-4.51     & 0.8  & 1.915  &1.915 &0.850   &0.8506  \\
0.04    & \-4.5429 &-4.5431   & 1.0  & 1.205  &1.205 &0.362   &0.3627  \\
0.08    &\-4.6775  &-4.6776   & 2.0  & 0.275  &0.275 &0.023   &0.0239  \\
0.10    &\-4.7843  &-4.7843   & 3.0  & 0.117  & --    &0.004   & --  \\
0.20    &\-5.9416  &-5.9416   & 4.0  & 0.064  & --    &0.001   & --  \\
0.43    &\-0.2971  & --    & 5.0  & 0.041  &0.041 &0.0005   &0.00057 \\
0.46    & 3.9273   & --    & 6.0  & 0.028  & --    &0.00027  & --  \\
0.465   &\-3.0867  & --    & 9.0  & 0.012  & --    &0.000049 & -- \\
0.477   & 1.2644   & --    & 10   &0.010081&0.01008&0.0000319 &0.00003\\
0.478   &\-1.9330  & --    &   &   &      &        &    \\
0.489   & \-0.6465 & --    &   &   &      &        &    \\
\end{tabular}
\end{ruledtabular}
\end{table*}
\begin{table*}
\caption{\label{tab:belo-khm}Kramers-Heisenberg matrix element $M$
in atomic units for $\omega$ below one photon ionization threshold.
Comparison is made with values in Ref. \cite{gav}.}
\begin{ruledtabular}
\begin{tabular}{llllcclc}
    &This     &     &  &This &    &         &This\\
 $\omega$(a.u)& Work &Ref. \cite{gav} &
 $\omega$(a.u) & Work &Ref. \cite{gav}  & $\omega$(a.u) & Work\\
 \hline %
 0.002  &\-0.000018&   --    &0.376 &  77.8416  & --    &      &        \\
 0.02   &\-0.0018  & \-0.0018  &0.38  &  15.3829  & 15.3828 &0.481 &4.0681      \\
 0.04   &\-0.0072  & \-0.0072  &0.4   &   2.6916  & 2.6916  &0.484 &0.6692      \\
 0.06   &\-0.0165  & \-0.0165  &0.429 &   0.0611  & --    &0.485 &\-0.4490    \\
 0.08   &\-0.0299  & \-0.0299  &0.43  & \-0.0549  &\-0.0549 &0.486 &\-16.087    \\
 0.10   &\-0.0478  & \-0.0478  &0.44  & \-3.1503  &\-3.1503 &        &      \\
 0.12   &\-0.0708  & \-0.0708  &0.444 &\-38.8927  & --    &0.488 &1.2367      \\
 0.14   &\-0.0999  & \-0.0999  &      &           &         &0.489 &\-0.1546    \\
 0.16   &\-0.1361  & \-0.1361  &0.445 &  32.2604  & 32.2603 &0.49  & 6.6498 \\
 0.18   &\-0.1812  & \-0.1812  &0.453 &  2.3124   & 2.3124  &0.491 & 1.2466 \\
 0.20   &\-0.2376  & \-0.2376  &0.464 &\-0.2004   & --    &0.492 &\-3.0681    \\
 0.22   &\-0.3091  & \-0.3091  &0.465 &\-0.6674   &\-0.6674 &0.493 & 1.2572 \\
 0.24   &\-0.4016  & \-0.4016  &0.468 &\-8.1693   &\-8.1693 &0.494 & 3.9947 \\
 0.26   &\-0.5246  & \-0.5246  &      &           &         &0.496 & 3.0200 \\
 0.30   &\-0.9507  & \-0.9507  &0.469 &  27.9814  &  27.9814&0.497 &\-3.1600    \\
 0.32   &\-1.3752  & \-1.3752  &0.473 &  1.97857  & 1.9785  &0.497 &\-3.1600    \\
 0.36   &\-5.3036  & \-5.3036  &0.477 &  0.2876   &--     &0.498 &\-0.7238    \\
 0.37   &\-15.763  & \-15.763  &0.478 &\-0.4416   &\-0.4416  &     &        \\
\end{tabular}
\end{ruledtabular}
\end{table*}
\begin{table*}
\caption{\label{tab:abo-khm}Real and Imaginary part of
Kramers-Heisenberg matrix element for photons of energies above one
photon ionization threshold $(\omega>0.5)$ and comparison is made
with values in Ref. \cite{gav}.}
\begin{ruledtabular}
\begin{tabular}{lllll}
&\multicolumn{2}{c}{Re $M$}&\multicolumn{2}{c}{Im $ M$}\\
$\omega$(a.u)&This work&Ref. \cite{gav}&This work&Ref. \cite{gav}   \\
 \hline
 0.6  & 1.1872  &1.1872  &0.9018  &0.9018  \\
 0.7  & 1.22161 &1.2216  &0.6900  &0.6900  \\
 0.8  & 1.22612 &1.2261  &0.5444  &0.5444  \\
 0.9  & 1.21842 &1.2184  &0.4400  &0.4400  \\
 1.0  & 1.20598 &1.2059  &0.3627  &0.3627  \\
 2.0  & 1.10007 &1.10007 &0.0958  &0.0958  \\
 3.0  & 1.05696 &1.0569  &0.0421  &0.0421  \\
 4.0  & 1.03685 &1.0368  &0.0231  &0.0231  \\
 5.0  & 1.02589 &1.0258  &0.0144  &0.0144  \\
 6.0  & 1.01924 &1.0192  &0.00977  &0.00977 \\
 7.0  & 1.01489 &1.0148  &0.00699  &0.00699 \\
 8.0  & 1.01188 &1.0118  &0.00522  &0.00522 \\
 9.0  & 1.00971 &1.0097  &0.00403  &0.00403 \\
 10  &  1.0081  &1.0081  &0.00319  &0.00319 \\
 20  &  1.00236 &1.0023  &0.00066  &0.00066 \\
 30  &  1.00112 &--    &0.000262 & --    \\
 40  &  1.00066 & --    &0.000133 & --    \\
 50  &  1.00042 & --    &0.000075 & --    \\
 90  &  1.00014 & --    &0.0000196& --   \\
\end{tabular}
\end{ruledtabular}
\end{table*}
\end{document}